\documentclass[%
 aip,
 amsmath,amssymb,
 reprint,%
]{revtex4-1}

\usepackage{graphicx}
\usepackage{dcolumn}
\usepackage{bm}
\usepackage{mathtools}
\usepackage[utf8]{inputenc}
\usepackage[T1]{fontenc}
\usepackage{color}
\usepackage{mathptmx}
\usepackage{hyperref}
\usepackage{soul}
\hypersetup{
	colorlinks=true,
	linkcolor=blue,
	citecolor=blue,
	filecolor=magenta,      
	urlcolor=blue,
}
\usepackage{siunitx}

\begin{document}

\title[Beckers \emph{et al.}\textemdash Generalized Boltzmann relations in semiconductors including band tails]{Generalized Boltzmann relations in semiconductors including band tails}

\author{Arnout Beckers}
\email{arnout.beckers@epfl.ch}
\affiliation{ICLAB, Ecole Polytechnique F\'ed\'erale de Lausanne (EPFL), 2000 Neuch\^atel, Switzerland}
\author{Dominique Beckers}
\affiliation{LRISK Research Center, Department of Insurance, KU Leuven, 3000 Leuven, Belgium}
\author{Farzan Jazaeri}
\affiliation{ICLAB, Ecole Polytechnique F\'ed\'erale de Lausanne (EPFL), 2000 Neuch\^atel, Switzerland}
\author{Bertrand Parvais}
\affiliation{Department of Electronics and Informatics,\,Vrije\,Universiteit Brussel,\,1050\,Brussels,\,Belgium}
\affiliation{Design-Enabled Technology Exploration Group, imec, 3001 Heverlee, Belgium}
\author{Christian Enz}
\affiliation{ICLAB, Ecole Polytechnique F\'ed\'erale de Lausanne (EPFL), 2000 Neuch\^atel, Switzerland}

\begin{abstract}
Boltzmann relations are widely used in semiconductor physics to express the charge-carrier densities as a function of the Fermi level and temperature. However, these simple exponential relations only apply to sharp band edges of the conduction and valence bands. In this article, we present a generalization of the Boltzmann relations accounting for exponential band tails. To this end, the required Fermi-Dirac integral is first recast as a Gauss hypergeometric function, followed by a suitable transformation of that special function, and a zeroth-order series expansion using the hypergeometric series. This results in simple relations for the electron and hole densities that each involve two exponentials. One exponential depends on the temperature and the other one on the band-tail parameter. The proposed relations tend to the Boltzmann relations if the band-tail parameters tend to zero. This work comes timely for the modeling of semiconductor devices at cryogenic temperatures for large-scale quantum computing. 
\end{abstract}

\maketitle

\section{Introduction}
Many phenomena in semiconductors cause an exponential decay of band states into the bandgap: e.g., extrinsic disorder due to dopants,\cite{kane_band_1985,halperin_impurity-band_1966,pankove_absorption_1965,mott} intrinsic static disorder due to native defects,\cite{teate_disorderinduced_1991,pichon_experimental_2011,soukoulis_exponential_1984} intrinsic dynamic disorder due to electron-phonon or electron-electron interactions,\cite{sarangapani} etc. The exponential \textquotedblleft band tail\textquotedblright \, is often referred to as a Lifshitz or Urbach tail,\cite{lifshitz_energy_1964,urbach,katahara_quasi-fermi_2014} and can serve to approximate the deeper tail portions of a Kane density-of-states (DOS) in heavily doped semiconductors \cite{kane,van_mieghem_theory_1992,sant_effect_2017} and a Gaussian DOS in organic semiconductors.\cite{haneef,li_compact_2010,horowitz_validity_2015} The decay length of the band tail (a.k.a. the band-tail parameter) is a measure of the disorder.\cite{kasap_springer_2017} Band tails have traditionally been linked to room-temperature (RT) applications with heavily disordered semiconductors, e.g., thin-film transistors, \cite{shur_physics_1984,hack_physical_1989,hernandez-barrios_analytical_2020} solar cells,\cite{boer,hack_theoretical_1983,poissant_analysis_2003,moore_importance_2016,nunomura_impact_2016} LEDs,\cite{okachi_determination_2009,bochkareva_efficiency_2013} lasers,\cite{stern_effect_1966,scifres_gaaspumped_1971} photodetectors,\cite{iacchetti_hopping_2012} and recently also tunnel FETs with high doping.\cite{khayer_effects_2011,agarwal_engineering_2014,memisevic_impact_2017} Cryogenic-temperature electronics is currently being revived for the classical control circuits of quantum bits and other quantum sensors.\cite{balestra_device_2001,charbon_cryo-cmos_2016,jeds,tracy_single_2016,tracy_integrated_2019} Recent studies\cite{edl_bohus,limit,ghibaudo_modelling_2020,hafez,cha} at cryogenic temperatures have revealed that band tails set a fundamental limit on the subthreshold performance and power consumption of MOSFETs and HEMTs fabricated from crystalline semiconductors with a relatively small amount of disorder near interfaces. This causes shallow band tails in the order of a few meVs, which have been measured using electron-spin resonance \cite{jock_probing_2012} and theoretically described using a path-integral approach.\cite{pinsook_description_2013} 

\section{Exponential Band Tails in Device Models}
Often responsible for sub-optimal device behavior, \cite{tiedje,gokmen_band_2013,bizindavyi_band-tails_2018,limit} band tails are an essential ingredient in analytical device models and simulation tools, and they need to be accounted for in the charge-carrier densities.\cite{shur_new_1989,hernandez-barrios_analytical_2020,tsormpatzoglou_analytical_2013,tedpaper} The required integral over an exponential DOS multiplied by the Fermi-Dirac (FD) function, cannot be solved analytically. The step-function (zero-Kelvin) approximation is usually assumed for the FD function to solve the integral.\cite{zhu,hack_physics_1985,lee_localized_2015,paasch_space-charge-limited_2009,hofacker,hormann_direct_2019} However, this approximation is only applicable if the band tail varies slowly over energy compared to the FD tail ($W_t=k_BT_c\gg k_BT$, where $k_BT$ is the thermal energy, $W_{t}$ the band-tail parameter,\footnote{Different notations are in use for the band-tail parameter: $E_0$ \cite{iribarren_band-tail_1998,khayer_effects_2011,scheinert_numerical_2014}, $E_U$ (Urbach) \cite{urbach,bochkareva_efficiency_2013}, $E_t$ \cite{zhu}, $\Delta E$ \cite{boer}, $W_0$\cite{herrmann_new_1993}, $W_t$ \cite{limit}, $k_BT_0$ \cite{scheinert_numerical_2014}, or $k_BT_t$ \cite{tsormpatzoglou_analytical_2013,servati_generalized_2006}. Here we opt for $W_t=k_BT_c$, in line with our earlier work \cite{limit}.} and $T_c$ the \textquotedblleft band-tail temperature\textquotedblright, i.e., the critical temperature at which the subthreshold swing in MOSFETs starts to saturate \cite{limit}).\cite{hack_physics_1985,hyun_kim_modeling_2011} So far, this approximation has been valid for most traditional RT applications with high disorder ($W_t\gg k_BT$), but it can lose its validity in the upcoming applications with smaller $W_t$ at lower temperatures ($W_t\approx\SI1-\SI{10}{\milli\electronvolt}, k_BT<\approx \SI{26}{\milli\electronvolt}$). 

Shur and Hack improved on the step-function approximation by proposing expression (8) in Ref.\cite{shur_physics_1984} valid for $W_t>k_BT$ [or $\alpha > 1$ in their notation, where $\alpha=W_t/(k_BT)$]. This is a popular expression in the literature.\cite{servati_generalized_2006,vissenberg,chen_physics-based_2018,paasch_space_2007} For the case $0<\alpha<1$, Shur and Hack introduced an interpolation function.\cite{shur_physics_1984} Paasch and Scheinert derived an expression for this case [see equation (A5) in Ref.\cite{paasch_space_2007}]. The lack of a generalization over $\alpha$ prevents deriving a device model that includes band tails and that is continuously valid across $T_c$ from room temperature (RT) down to the deep-cryogenic regime.\cite{tedpaper} 

It is a lesser known fact that the required FD integral including exponential DOS can be recast as a Gauss hypergeometric function of the form $H_{2F1}(1,\theta;\theta+1;z)$ \cite{limit}, where $\theta=1/\alpha$ and $z$ is the main argument. This special function is valid for any $T$, $W_t$, and position of the Fermi level ($E_F$), and has some useful properties, see \cite{rhyzik,nist,oldham_atlas_2009} and Appendix \ref{app:gauss}. $H_{2F1}$ itself is not convenient for semiconductor device analysis because it is typically not available in standard tools. Furthermore, it cannot\footnote{because one of the first two parameters in $H_{2F1}$ equals 1, leading to a division-by-zero error.} be integrated with formula (\ref{eq:int}) presented in Appendix \ref{app:gauss}. The aim of this study is to apply suitable properties to $H_{2F1}$ to arrive at an analytical expression that unifies the expression from Shur and Hack, the step-function approximation, and the high-temperature limit. Section \ref{sec:deriv} derives such a generalized expression for a carrier density in a band tail. Section \ref{sec:complete} adds the band carriers and makes the link to the Boltzmann relations for the electron and hole densities when the band tails reduce to sharp band edges.
\section{\label{sec:deriv}Generalized Expression of a Carrier Density in a Band Tail}
\begin{figure}[t]
	\centering
	\includegraphics[width=0.45\textwidth]{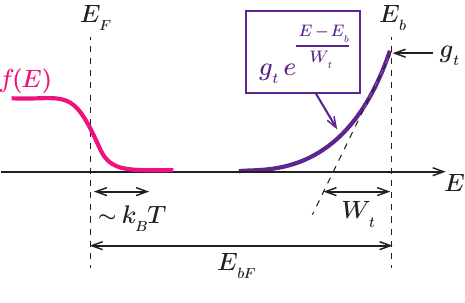}
	\vspace{-0.5cm}
	\caption{Exponential band-tail DOS (purple) and FD function $f(E)$ (pink). The band tail has a characteristic decay $W_t$ in the bandgap.}
	\label{fig:int}
\end{figure}

\begin{figure*}[t]
	\centering
	\includegraphics[width=0.95\textwidth]{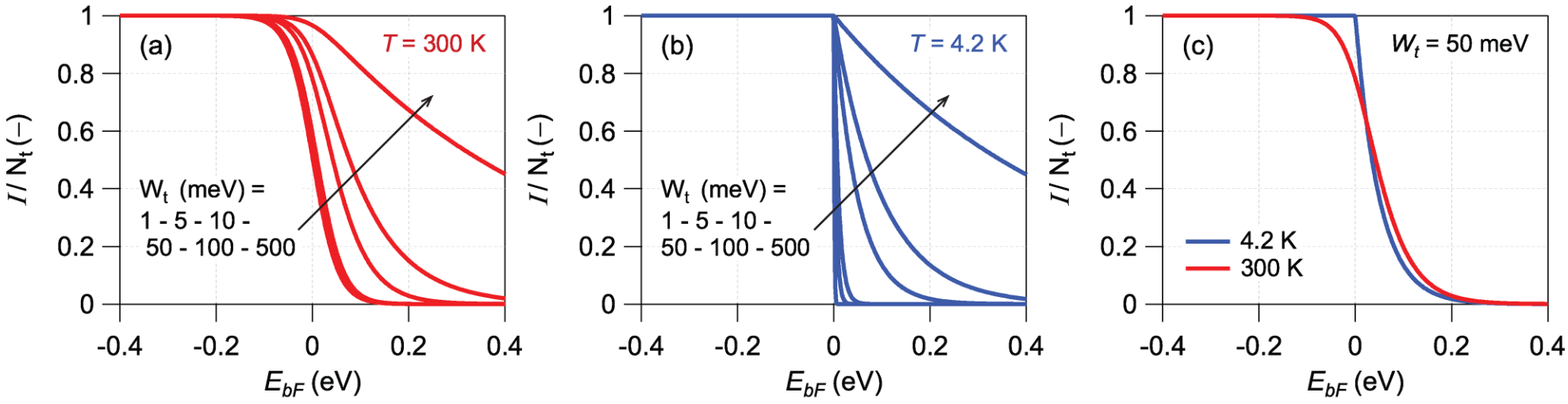}
	\caption{Expression (\ref{eq:nbt2}). The Gauss hypergeometric function $H_{2F1}(1,\theta;\theta+1;z)$, where $z=-\exp(E_{bF}/k_BT)$ and $\theta=k_BT/W_t$, is plotted (a) at $T=\SI{300}{\kelvin}$, (b) at $T=\SI{4.2}{\kelvin}$ for different $W_t$, and (c) at $W_t=\SI{50}{\milli\electronvolt}$. In this article, we derive an approximation for $H_{2F1}$ for positive $E_{bF}$.}
	\label{fig:hyptif}
\end{figure*}

As shown in Fig.\,\ref{fig:int}, the integral over the multiplication of the exponential band tail and the FD function is given by 
\begin{equation}
I =g_t\int_{-\infty}^{E_b} \exp\left(\frac{E-E_b}{W_t}\right)f(E)dE,
\label{eq:int1}
\end{equation}
where $g_t$ is the value of the DOS ($\si{\per\centi\meter\cubed\per\electronvolt}$) at the band edge ($E_b$), $W_t$ the band-tail parameter, and $f(E)=1/\left[1+\exp\{ \left(E-E_F\right)/(k_BT)\}\right]$ the FD function. Integral $I$ can represent either an electron density in a conduction-band tail $(n_{bt})$ or a hole density in a valence-band tail ($p_{bt}$). The complete electron and hole density relations with a joint DOS (including the band carriers that are not shown in Fig.\,\ref{fig:int}) will be derived in Section \ref{sec:complete}. Using the substitution $z=\exp\left[(E-E_F)/W_t\right]$, integral (\ref{eq:int1}) can be rewritten as:\cite{shur_physics_1984} 
\begin{equation}
I=N_t\exp\left(\frac{-E_{bF}}{W_t}\right)J(\alpha),
\label{eq:nbt}
\end{equation}
where $N_t=g_tW_t$ is the total density of states ($\si{\per\centi\meter\cubed}$) in the band tail, $E_{bF}=E_b-E_F$, and 
\begin{equation}
J(\alpha)=\int_0^{\exp\left(\frac{E_{bF}}{W_t}\right)}\frac{dz}{1+z^\alpha}, 
\end{equation}
where $\alpha=W_t/(k_BT)$ is the dispersion parameter.\cite{kasap_springer_2017} Using (\ref{eq:fd}) from Appendix \ref{app:gauss}, integral $J(\alpha)$ can be recast as: 
\begin{equation}
J(\alpha)=\left[zH_{2F1}\left(1,\frac{1}{\alpha};1+\frac{1}{\alpha};-z^\alpha\right)\right]_0^{\exp\left(\frac{E_{bF}}{W_t}\right)}
\label{eq:JalphaRecast}
\end{equation}
where $H_{2F1}$ is the Gauss hypergeometric function\cite{rhyzik}, see Appendix \ref{app:gauss} for more 
details about this function. Combining (\ref{eq:JalphaRecast}) with (\ref{eq:nbt}), we obtain
\begin{equation}
\frac{I}{N_t}=H_{2F1}\left(1,\theta;\theta+1;z\right), 
\label{eq:nbt2}
\end{equation} 
where $\theta$ is defined as the inverse dispersion parameter ($\theta=1/\alpha=k_BT/W_t=T/T_c$), and $z=-\exp\left(E_{bF}/k_BT\right)$. Expression (\ref{eq:nbt2}) is valid for all positions of $E_F$ wrt $E_b$ and for all $T$ wrt $T_c$ (all $\alpha$ or $\theta$). It is plotted in Fig.\,\ref{fig:hyptif} for different $T$ and $W_t$. The Gauss hypergeometric function resembles an asymmetric sigmoid function,  which levels off quicker or slower depending on $T$ and $W_t$. For a position of $E_F$ in the band ($E_{bF}< 0$), expression (\ref{eq:nbt2}) simply reduces\footnote{$I\approx N_tH_{2F1}(1,\theta;\theta+1;0)$, and use the special value (\ref{eq:specialvalue}) in Appendix \ref{app:gauss}} to $I\approx N_t$, as is clear from Figs.\,\ref{fig:hyptif}(a)-(c). 

$H_{2F1}$ can be expressed as the hypergeometric series, given by (\ref{eq:pow}) in Appendix \ref{app:gauss}. This is the only series known for $H_{2F1}$.\cite{rhyzik,nist,oldham_atlas_2009} This series converges only if $\vert z \vert < 1$,\cite{rhyzik} which means in our case that $E_{bF}<0$, or only applicable for a position of $E_F$ in the band, for which we already know the solution. Transformation formulae \cite{rhyzik,nist,oldham_atlas_2009} can be applied to $H_{2F1}$ to allow a series expansion of $H_{2F1}$ when $E_{bF}>0$. However, that expression will only be valid if $E_F$ lies in the tail states. Thus, the convergence criterion of the hypergeometric series ($\vert z\vert<1$) prevents us from simplifying $H_{2F1}$ into an analytic expression that is valid for all positions of $E_F$ with respect to the band edge. The remainder of this paper  will focus on the case $E_{bF}>0$. 

\subsection{\label{sec:lco}Transformation of the Gauss hypergeometric function}
\begin{figure}[t]
	\centering
	\includegraphics[width=0.49\textwidth]{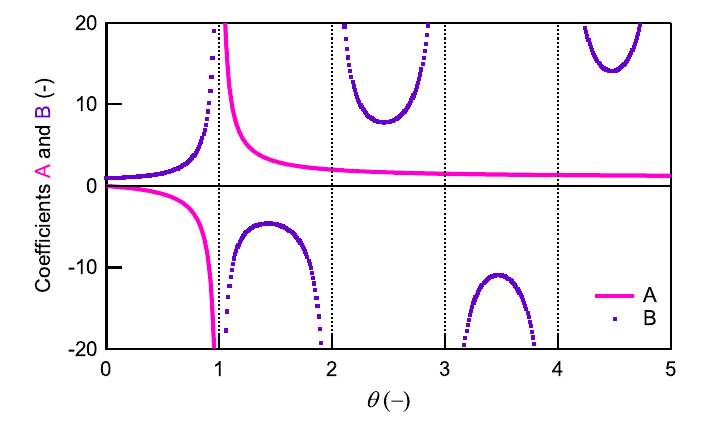}
	\vspace{-0.75cm}
	\caption{Coefficients $A=\theta/(\theta-1)$ and $B=1/\mathrm{sinc}(\pi\theta)$ from expressions (\ref{eq:AB}), (\ref{eq:lco3}), (\ref{eq:lco2}) and (\ref{eq:zerothorder}) versus $\theta\triangleq k_BT/W_t$.}
	\label{fig:N=0}
\end{figure}

Transformations can be applied to Gauss hypergeometric functions to bring their argument $z$ within the unit circle of convergence ($\vert z \vert < 1$).\cite{rhyzik} The series expansion (\ref{eq:pow}) can then be carried out. Two suitable transformations are the Pfaff transformation and the linear-combination transformation [see (\ref{eq:app-pfaff}) and (\ref{eq:app-lco}) in Appendix \ref{app:gauss}]. The simplest one, the Pfaff transformation, has too slow convergence to be practically useful, as discussed in Appendix \ref{sec:pfaff}. Therefore, we have to use the more elaborate (\ref{eq:app-lco}), which transforms $z$ to $\zeta=1/(1-z)$. Since $\vert \zeta\vert\ll1$ if $E_{bF}>0$, rapid convergence can be expected in the hypergeometric series. The disadvantage is that it introduces Gamma functions, which have discontinuities on the negative axis.\cite{rhyzik} An alternative transformation that does not introduce discontinuities was not found in the standard references\cite{rhyzik,nist,oldham_atlas_2009} nor online.\cite{wolfram} Applying thus transformation (\ref{eq:app-lco}) to $H_{2F1}$ in (\ref{eq:nbt2}) and using the Gamma-function identities $\Gamma(\theta+1)=\theta\Gamma(\theta)$ and $\Gamma(\theta)=(\theta-1)\Gamma(\theta-1)$, gives 
\begin{eqnarray}
H_{2F1}\left(1,\theta;\theta+1;z\right)&=&\zeta\frac{\theta}{\theta-1}H_{2F1}(1,1;2-\theta;\zeta)\\
&+&\zeta^\theta\theta\Gamma (\theta)\Gamma(1-\theta)H_{2F1}(\theta,\theta;\theta;\zeta),\nonumber
\label{eq:lco1}
\end{eqnarray}
where $\zeta=1/(1-z)$ and $z=-\exp\left[E_{bF}/(k_BT)\right]$. The Gamma functions are not of closed-form. Using Euler's reflection formula,\cite{rhyzik} 
\begin{equation}
\Gamma(\theta)\Gamma(1-\theta)=\frac{\pi}{\sin(\pi \theta)}, 
\end{equation}
we find that $\theta\Gamma(\theta)\Gamma(1-\theta)=\pi\theta/\sin(\pi\theta)=1/\mathrm{sinc}(\pi\theta)$, where the sine cardinal ($\mathrm{sinc}$) function
appears, which is also present in the popular case $\alpha>1$ of Shur and Hack.\cite{shur_physics_1984} Due to the $\mathrm{sinc}$ function, expression (\ref{eq:lco1}) is only valid if $\theta \notin \mathbb{N}_0$, and thus discontinuities will arise when the temperature equals integer multiples of the characteristic band-tail temperature $T_c$, i.e., $k_BT= mW_t$ or $T= mT_c$ with $m\in \mathbb{N}_0$. Defining the coefficients
\begin{subequations}
\label{eq:AB}
\begin{eqnarray}
A&=&\frac{\theta}{\theta-1},\\
B&=&\frac{1}{\mathrm{sinc}(\pi\theta)},
\end{eqnarray}
\end{subequations}
we obtain
\begin{eqnarray}
H_{2F1}\left(1,\theta;\theta+1;z\right)&=A\zeta H_{2F1}(1,1;2-\theta;\zeta)\nonumber\\
&+B\zeta^\theta H_{2F1}(\theta,\theta;\theta;\zeta).
\label{eq:lco3}
\end{eqnarray}
$A$ and $B$ are plotted vs. $\theta$ in Fig.\ref{fig:N=0}. $A$ has a vertical asymptote in $\theta=1$ and $B$ in $\theta\in  \mathbb{N}_0$. The sinc-function in $B$ occurs often in the context of band tails\cite{shur_physics_1984,marinov_compact_2009,kopprio_obtainment_2017} but it is a source of discontinuities.

\subsection{Hypergeometric series expansion}
\begin{figure*}[ht]
	\centering
	\includegraphics[width=0.95\textwidth]{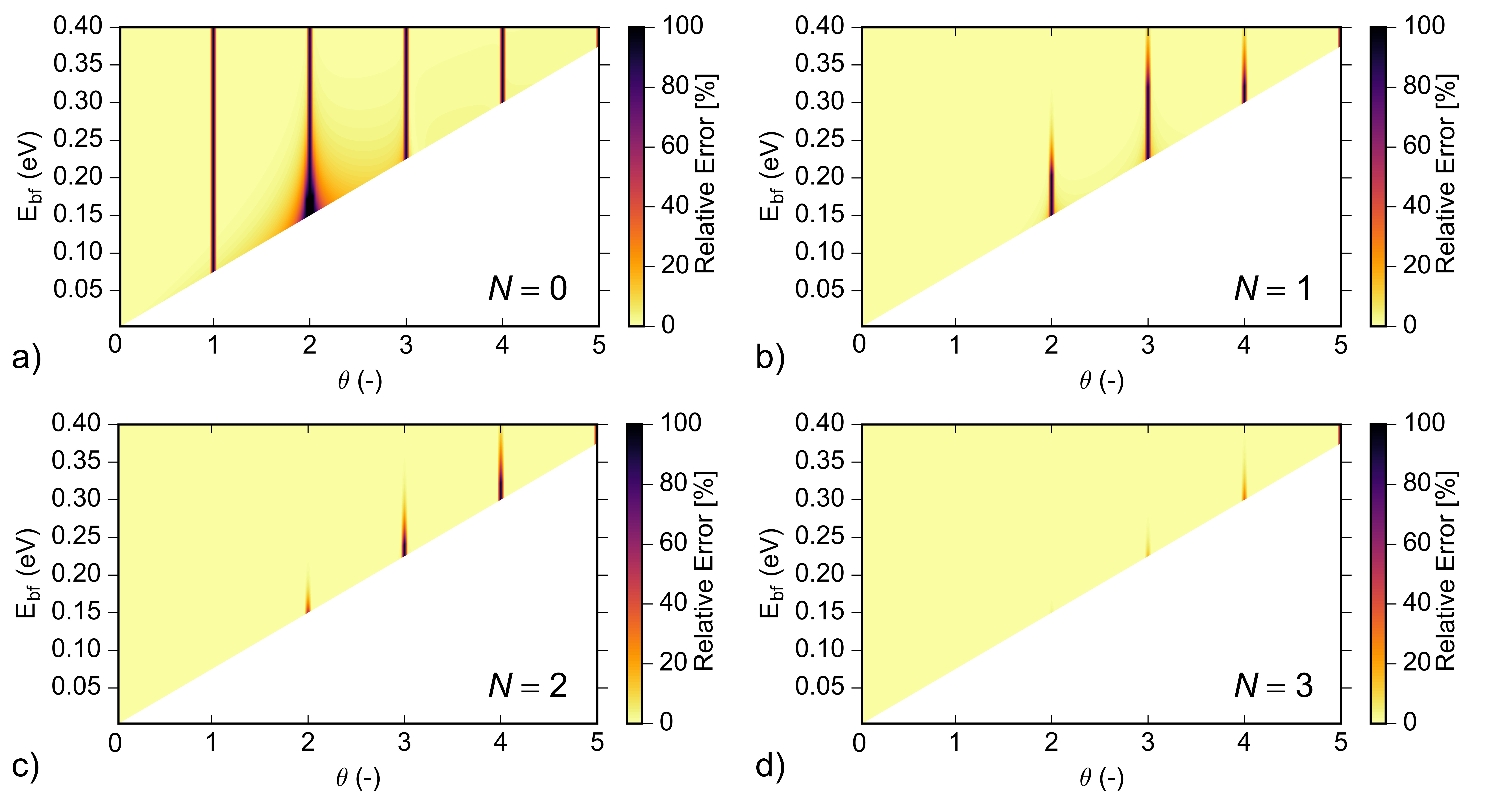}
	\vspace{-0.5cm}
	\caption{a) Relative percentage error between (\ref{eq:zerothorder}) and (\ref{eq:nbt2}) for $E_{bf}>3k_BT$. Error peaks are visible near the discontinuities $\theta\in \mathbb{N}_0$. b) to d) Higher order approximations of (\ref{eq:lco3}) decrease the error around the discontinuities. ($\theta\triangleq k_BT/W_t$, $\theta$-step: 0.03).}
	\label{fig:N=0-1-2}
	\vspace*{\floatsep}
	
	\includegraphics[width=0.95\textwidth]{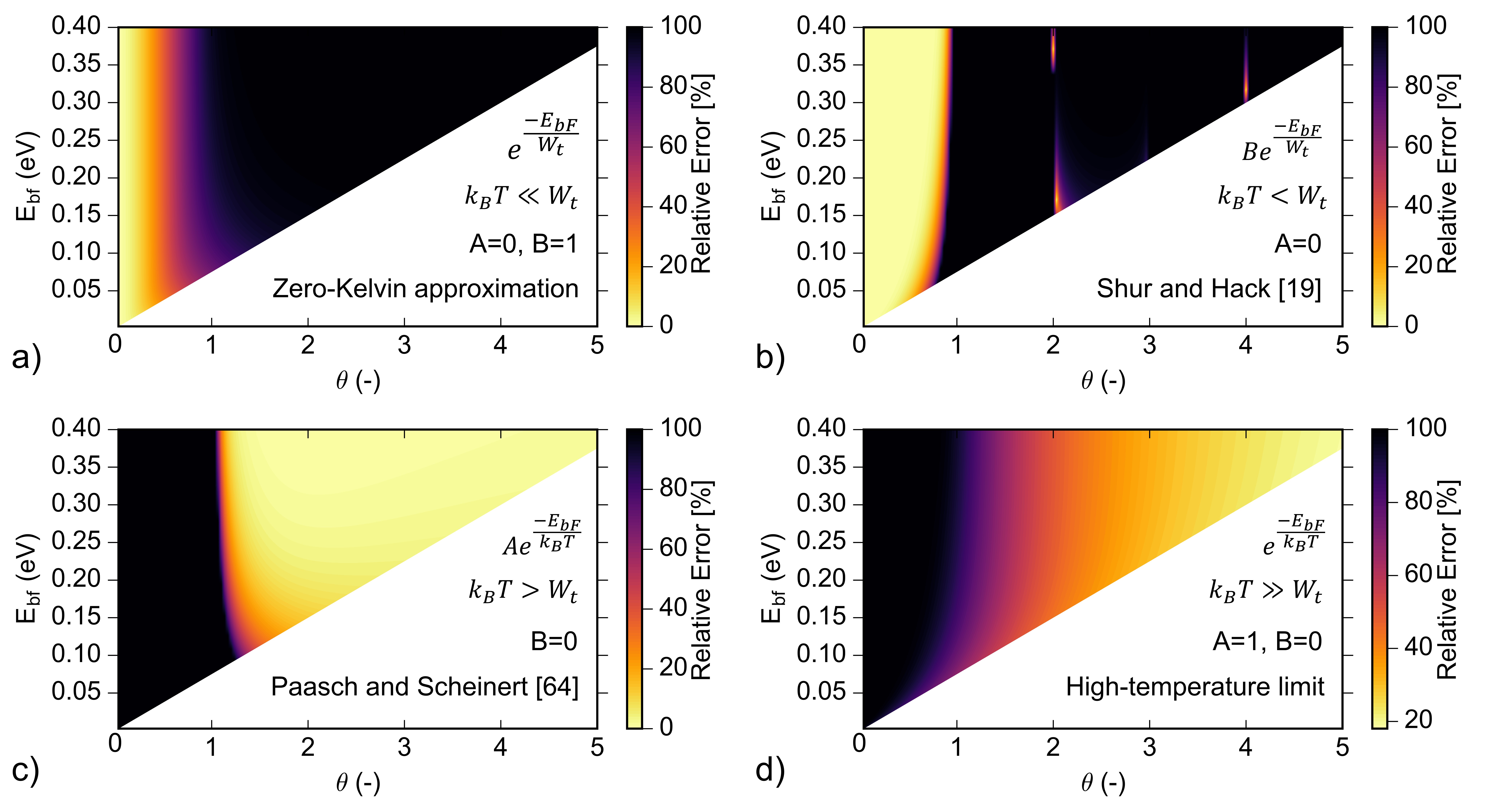}
	\vspace{-0.5cm}
	\caption{Relative percentage error between four different approximations of (\ref{eq:zerothorder}) ($N=0$) and the exact (\ref{eq:nbt2}). Expression (\ref{eq:zerothorder}) unifies these four different  cases, a) zero-Kelvin approximation,\cite{zhu,hack_physics_1985,lee_localized_2015,paasch_space-charge-limited_2009,hofacker,hormann_direct_2019} b) Shur and Hack [expression (8) in Ref.\cite{shur_physics_1984}], c) see equation (A5) in Ref.\cite{paasch_space_2007}, d) high-temperature/low-disorder limit. [$A=\theta/(\theta-1)$ and $B=1/\mathrm{sinc}(\pi\theta)$, where $\theta\triangleq k_BT/W_t$].}
	\label{fig:N=0_cases}
\end{figure*}

Expanding both $H_{2F1}$-functions in (\ref{eq:lco3}) into their hypergeometric series (\ref{eq:pow}) given in Appendix \ref{app:gauss}, yields
\begin{subequations}
	\begin{eqnarray}
	\label{eq:lco2}H_{2F1}\left(1,\theta;\theta+1;z\right)&=& A\zeta S_1+ B\zeta^\theta S_2,\\
	\label{eq:S1}S_1&=&\sum_{j=0}^{N} \frac{(1)_j(1)_j}{(2-\theta)_j}\frac{\zeta^j}{j!},\\
	\label{eq:S2}S_2&=&\sum_{j=0}^{N}\frac{(\theta)_j\zeta^j}{j!},
	\end{eqnarray}
\end{subequations}
where $\zeta=1/(1-z)$, and $z=-\exp\left[E_{bF}/(k_BT)\right]$. Since $E_{bF}> \,\approx3k_BT$, the Maxwell-Boltzmann approximation can be applied to $\zeta$, leading to $\zeta\approx \exp(-E_{bF}/k_BT)$. Taking only the first term of the series ($N=0$) in (\ref{eq:S1}) and (\ref{eq:S2}), gives 
\begin{equation}
H_{2F1}\left(1,\theta;\theta+1;z\!\right)\!\approx A\exp\left(\frac{-E_{bF}}{k_BT}\!\right)\!+B\exp\left(\frac{-E_{bF}}{W_t}\!\right)
\label{eq:zerothorder}
\end{equation}
Expressions (\ref{eq:lco2}) and (\ref{eq:zerothorder}) cannot be evaluated exactly at the discontinuities $\theta \in \mathbb{N}_0$ due to $A$ and $B$. Figure \ref{fig:N=0-1-2}(a) shows the relative percentage error between the LHS and the RHS of (\ref{eq:zerothorder}). Note that the contour plot shows the error in close proximity to the discontinuities, not exactly at the discontinuities. Far enough from a discontinuity, expression (\ref{eq:zerothorder}) is an excellent approximation of $H_{2F1}$. 

The error around the discontinuities gradually disappears when using (\ref{eq:lco2}) with a higher $N$ than $N=0$, as shown in Figs.\,\ref{fig:N=0-1-2}(b)-(d), starting around $\theta=1$, $\theta=2$, and so on. The error around the discontinuities is thus due to truncating the series at $N=0$. The higher-order approximations for ($N=1$) and ($N=2$) are written down in Appendix \ref{sec:higher}. The discontinuities at $\theta\in \mathbb{N}_0$ are in principle removable, because the values at $\theta\in \mathbb{N}_0$ can be directly computed from integral (\ref{eq:int1}) and a compound function could be defined.

Figures \ref{fig:N=0_cases}(a)-(d) show that (\ref{eq:zerothorder}) unifies the zero-Kelvin (step-function) approximation ($k_BT\ll W_t$) [Fig. \ref{fig:N=0_cases}(a)], the expression of Shur and Hack\cite{shur_physics_1984} for ($\alpha>1$) or ($k_BT< W_t$) [Fig. \ref{fig:N=0_cases} (b)], and the high-temperature cases ($k_BT>W_t$) [Fig. \ref{fig:N=0_cases}(c)] and ($k_BT\gg W_t$) [Fig. \ref{fig:N=0_cases}(d)], see e.g., the Appendix in Ref.\cite{paasch_space-charge-limited_2009}. Despite its discontinuities at $T=mT_c$, approximation (\ref{eq:zerothorder}) is still an improvement over the four different approximations in Fig.\,\ref{fig:N=0-1-2}, which each show large regions where the error is intolerable. In the next section, we will use approximation (\ref{eq:zerothorder}) ($N=0$) for the band-tail carrier densities ($n_{bt}$ and $p_{bt}$) to derive simple relations for the total carrier densities (including both the band-tail and the band carriers). 
 \begin{figure*}[t]
	\centering
	\includegraphics[width=0.9\textwidth]{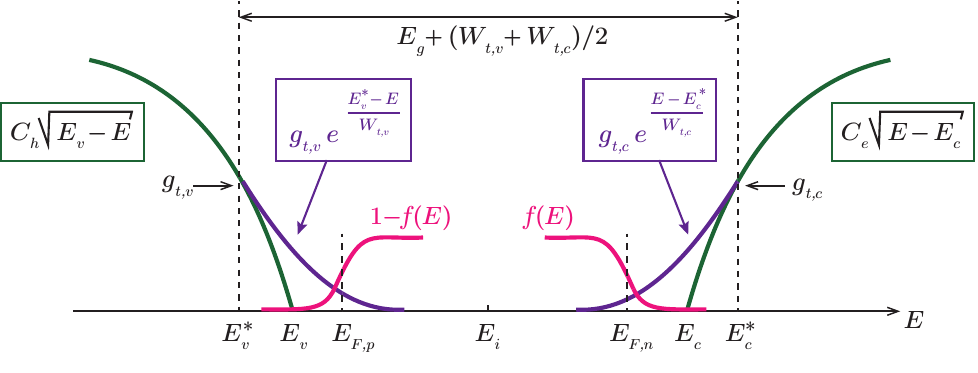}
	\vspace{-0.5cm}
	\caption{Conduction band and valence band DOS (green) including exponential band tails (purple). The density-of-state tails have characteristic decays of $W_{\mathrm{t,c}}$ and $W_{\mathrm{t,v}}$, respectively. Imposing continuity conditions at $E_{c/v}^*$, it follows that $E_{c/v}^*=E_{c/v}\pm W_{t,c/v}/2$.\cite{wager}}
	\label{fig:3D_DOS}
\end{figure*}
\section{\label{sec:complete}Relations Including Band Carriers}
\subsection{In terms of effective DOS ($N_c, N_v$)}
As shown in Fig.\,\ref{fig:3D_DOS}, the exponential band tail from Fig.\,\ref{fig:int} can be joined to the bulk DOS of the conduction and valence bands. Here, $E_{c/v}^*$ take the role of $E_b$ from the previous section. This \textquotedblleft joint DOS\textquotedblright \, is a commonly used approach\cite{jiao_improved_1998,oleary_simplified_2002,wager} to approximate the Kane DOS,\cite{kane_band_1985,van_mieghem_theory_1992} which is not convenient for semiconductor-device analysis because it is given by a parabolic cylinder function. Using a joint DOS allows to split the FD integral in two separate integrals of band-tail carriers and band carriers. We have assumed a 3-D (bulk) DOS for the band carriers, but our derivation can also apply to other types of band DOS with an exponential tail. For the holes, we find $p=p_b+p_{bt}$, where 
\begin{subequations}
\begin{eqnarray}
p_b&=&\int_{-\infty}^{E^*_{v}}C_h\sqrt{E_v-E}\left[1-f(E)\right]dE,
\label{eq:pb}\\
\label{eq:pbtx}p_{bt}&=&\int_{E^{*}_v}^{\infty}g_{t,v}\exp\left(\frac{E_v^*-E}{W_{t,v}}\right)\left[1-f(E)\right]dE. 
\end{eqnarray}
\end{subequations}
For the electrons, we have that $n=n_{bt}+n_{b}$, where 
\begin{subequations}
\begin{eqnarray}
\label{eq:nbtx}n_{bt}&=&\int_{-\infty}^{E^*_c}g_{t,c}\exp\left(\frac{E-E_c^*}{W_{t,c}}\right)f(E)dE,\\
n_b&=&\int_{E^*_c}^{\infty}C_e\sqrt{E-E_c}f(E)dE.\label{eq:nb}
\end{eqnarray}
\end{subequations}
The coefficients $C_{e/h}$ in (\ref{eq:pb}) and (\ref{eq:nb}) are given by $C_{e/h}=m_{\mathrm{de/dh}}\sqrt{2m_{\mathrm{de/dh}}}/(\hbar^3\pi^2)$ ($m_{\mathrm{de/dh}}$ are the electron and hole density-of-state effective masses).

Integrals (\ref{eq:pb}) and (\ref{eq:nb}) for the band carriers $p_b$ and $n_b$ are standard FD integrals of order 1/2, except that their integration starts or ends at $E_{c/v}^*$ instead of the usual $E_{c/v}$. The solutions of (\ref{eq:pb}) and (\ref{eq:nb}) are given in Appendix \ref{sec:app}. Integrals (\ref{eq:pbtx}) and (\ref{eq:nbtx}) for the band-tail carriers $p_{bt}$ and $n_{bt}$ can be rewritten as a Gauss hypergeometric function using (\ref{eq:nbt2}) from the previous section. We obtain $p_{bt}=N_{\mathrm{t,v}}H_{2F1}\left(1,\theta_\mathrm{v};\theta_\mathrm{v}+1;z_\mathrm{v}\right)$ and $n_{bt}=N_{\mathrm{t,c}}H_{2F1}\left(1,\theta_\mathrm{c};\theta_\mathrm{c}+\!1;z_\mathrm{c}\right)$, where 
$\theta_\mathrm{v}=k_BT/W_{\mathrm{t,v}}$, $z_\mathrm{v}=-\exp\left[(E_{F,p}-E^*_v)/(k_BT)\right]$, $\theta_\mathrm{c}=k_BT/W_{\mathrm{t,c}}$, and $z_\mathrm{c}=-\exp\left[(E^*_c-E_{F,n})/(k_BT)\right]$.  Using approximation (\ref{eq:zerothorder}) on $p_{bt}$ and $n_{bt}$, we find that $p\approx p_{bt,1}+p_{bt,2}+p_b$, where
\begin{subequations}
	\begin{eqnarray}
	\label{eq:pbt1_final} p_{bt,1}&=& N_{\mathrm{t,v}}A_\mathrm{v}\exp\left(\frac{E_v^*-E_{F,p}}{k_BT}\right)\\
	\label{eq:pbt2_final}p_{bt,2}&=&N_{\mathrm{t,v}}B_\mathrm{v}\exp\left(\frac{E_v^*-E_{F,p}}{W_{\mathrm{t,v}}}\right)\\
	\label{eq:pb_final}p_{b}&=&N_\mathrm{v}^\prime\exp\left(\frac{E_v-E_{F,p}}{k_BT}\right)
	\end{eqnarray}
\end{subequations}
and that $n\approx n_{bt,1}+n_{bt,2}+n_b$, where 
\begin{subequations}
	\begin{eqnarray}
	\label{eq:nbt1_final}n_{bt,1}&=&N_{\mathrm{t,c}}A_\mathrm{c}\exp\left(\frac{E_{F,n}-E_c^*}{k_BT}\right)\\
	\label{eq:nbt2_final}n_{bt,2}&=&N_{\mathrm{t,c}}B_\mathrm{c}\exp\left(\frac{E_{F,n}-E_c^*}{W_{\mathrm{t,c}}}\right)\\
	\label{eq:nb_final}n_{b}&=&N_c^\mathrm{\prime}\exp\left(\frac{E_{F,n}-E_c}{k_BT}\right)
	\end{eqnarray}
\end{subequations}
Expressions (\ref{eq:pbt1_final})-(\ref{eq:pb_final}) and (\ref{eq:nbt1_final})-(\ref{eq:nb_final}) are valid for $E_{F,p} > E_{v}^*+3k_BT$ and $E_{F,n}< E_c^*-3k_BT$, respectively. Note that we have used the notation $A_v=A(\theta_\mathrm{v})$, $B_c=B(\theta_\mathrm{c})$, etc. The band carriers $n_b$ and $p_b$ are similar to the Boltzmann relations, but have a lower effective density of states given by 
\begin{equation}
N_{\mathrm{c/v}}^\prime=N_{\mathrm{c/v}}\left[1-\frac{2}{\sqrt{\pi}}\gamma\left(\frac{3}{2},\frac{1}{2\theta_{\mathrm{c/v}}}\right)\right],
\end{equation}
which has been derived in Appendix \ref{sec:app}. The values of $N_{t,c/v}$ and $E_{c/v}^*$ can be determined by imposing continuity conditions at $E_{c/v}^*$.\cite{wager} They are satisfied if $E_v-E_v^*=W_{\mathrm{t,v}}/2$, $E^*_c-E_c=W_{\mathrm{t,c}}/2$, and $N_{\mathrm{t,c/v}}=C_{e/h}W_\mathrm{{t,c/v}}\sqrt{0.5W_\mathrm{{t,c/v}}}$.
\subsection{In terms of the intrinsic carrier density ($n_i$)}
Boltzmann relations are usually given in terms of the intrinsic carrier density $n_i$. For an intrinsic semiconductor ($E_{F,n}=E_{F,p}=E_i$) with band tails as shown in Fig.\,\ref{fig:3D_DOS}, three separate mass-action laws can be written: (1) $n_{bt,1}\cdot p_{bt,1}=n_{i,bt1}^2, (2) \, n_{bt,2}\cdot p_{bt,2}=n_{i,bt2}^2$, and (3) $n_b\cdot p_b=n_{i,b}^2$. Combining these with (\ref{eq:pbt1_final})-(\ref{eq:pb_final}) and (\ref{eq:nbt1_final})-(\ref{eq:nb_final}), we obtain the following three components of $n_i=n_{i,bt1}+n_{i,bt2}+n_{i,b}$: 
\begin{eqnarray}
\label{eq:nibt1}n_{i,bt1}&=&C_1\exp\left(\frac{-E_g}{2k_BT}\right)\exp\left(\!\frac{-W_{t,c}-W_{t,v}}{4k_BT}\!\right)\\
\label{eq:nibt2}n_{i,bt2}&=&C_2\exp\left(\frac{-E_g}{4W_{t,c}}\right)\exp\left(\frac{-E_g}{4W_{t,v}}\right)e^{-0.5}\\
\label{eq:nib}n_{i,b}&=&\sqrt{N_c^\prime N_v^\prime}\exp\left(\frac{-E_g}{2k_BT}\right)
\end{eqnarray}
where $C_1=\sqrt{N_{t,c}N_{t,v}A_cA_v}$ and $C_2=\sqrt{N_{t,c}N_{t,v}B_cB_v}$. Notice that expression (\ref{eq:nib}) is similar to the standard $n_i$. Finally, we obtain the generalized Boltzmann relations for electron and hole densities including exponential band tails: 
\begin{subequations}
	\begin{eqnarray}
	\label{eq:boltzn}n\approx (n_{i,bt1}&+&n_{i,b})\exp\left(\frac{E_{F,n}-E_i}{k_BT}\right)\\&+&n_{i,bt2}\exp\left(\frac{E_{F,n}-E_i}{W_{t,c}}\right)\nonumber\\
	\label{eq:boltzp}p\approx(n_{i,bt1}&+&n_{i,b})\exp\left(\frac{E_i-E_{F,p}}{k_BT}\right)\\&+&n_{i,bt2}\exp\left(\frac{E_i-E_{F,p}}{W_{t,v}}\right)\nonumber
	\end{eqnarray}
\end{subequations}
\begin{figure}[t]
	\includegraphics[width=0.45\textwidth]{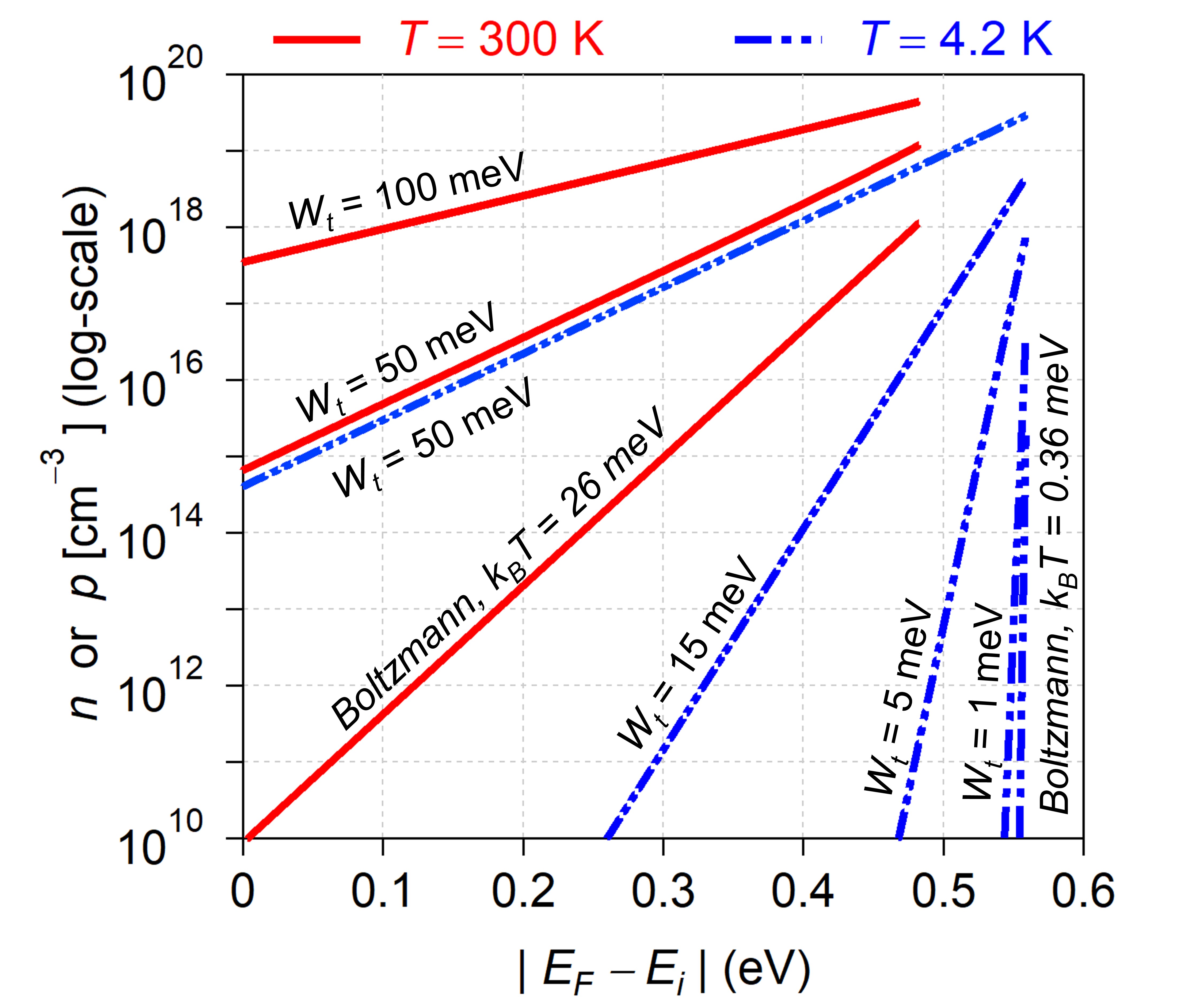}
	\caption{New relations (\ref{eq:boltzn})-(\ref{eq:boltzp}) including band tails (annotated with a $W_t$) versus the standard Boltzmann relations (annotated with \textquotedblleft Boltzmann\textquotedblright) at \SI{300}{\kelvin} and $\SI{4.2}{\kelvin}$ for $\vert E_F-E_i\vert<E_g/2-3k_BT$. The bandgap of silicon is assumed. If $W_\mathrm{t}<k_BT$, the new relations fall back on the Boltzmann relations.}
	\label{fig:nrelations}
\end{figure}
We can distinguish between the following limits of (\ref{eq:boltzn}) and (\ref{eq:boltzp}): 
\begin{itemize}
	\item In the limit of sharp band edges ($W_\mathrm{t,c/v}\rightarrow 0$), we find that $N_\mathrm{t,c/v}\rightarrow 0$ and thus $n_{i,bt1}$ and $n_{i,bt2}\rightarrow 0$. Furthermore, we have that $n_{i,b}\rightarrow n_{i}$, because $N_{c/v}^\prime \rightarrow N_{c/v}$ when $\theta_{c/v} \rightarrow \infty$. Thus (\ref{eq:boltzn}) and (\ref{eq:boltzp}) fall back on the standard Boltzmann relations, $n=n_i\exp[(E_{F,n}-E_i)/(k_BT)]$ and $p=n_i\exp[(E_{i}-E_{F,p})/(k_BT)]$, as shown in Fig.\,\ref{fig:nrelations}. 
	\item In the limit of high temperatures and/or low disorder ($k_BT\gg W_\mathrm{t,c/v}$), we have that $A \rightarrow 1$, $B\rightarrow 0$, and $n_{i,bt2}\rightarrow 0$. We obtain $n\approx (n_{i,bt1}+n_{i,b})\exp\left[(E_{F,n}-E_\mathrm{i})/(k_BT)\right]$ and $p\approx (n_{i,bt1}+n_{i,b})\exp\left[(E_\mathrm{i}-E_{F,p})/(k_BT)\right]$. At sufficiently high temperature, the band-tail states will become negligible compared to the band states ($n_{i,bt1}\ll n_{i,b}$), returning to the standard Boltzmann relations. 
	\item In the limit of low temperatures and/or high disorder $(k_BT\ll W_\mathrm{t,c/v})$, we have that $A \rightarrow 0$, $B\rightarrow 1$, and $n_{i,bt1}\rightarrow 0$. Since $n_{i,b}\ll n_{i,bt2}$, we obtain $n\approx n_{i,bt2}\exp\left[(E_{F,n}-E_\mathrm{i})/W_{t,c}\right]$ and $p\approx n_{i,bt2}\exp\left[(E_\mathrm{i}-E_{F,p})/W_{t,v}\right]$. Expanding $n_{i,bt2}$, it can be seen that these expressions are equal to solutions of (\ref{eq:int1}) when assuming the step-function approximation for $f(E)$: $n\approx N_{\mathrm{t,c}}\exp[\left(E_{F,n}-E_c^*\right)/W_{\mathrm{t,c}}]$ and $p\approx N_{\mathrm{t,v}}\exp[\left(E_v^*-E_{F,p}\right)/W_{\mathrm{t,v}}]$. 
\end{itemize}
As shown in Fig.\,\ref{fig:nrelations}, if $k_BT\ll W_\mathrm{t,c/v}$, the slope of the charge-carrier density is inversely proportional to $W_\mathrm{t,c/v}$. On the other hand, if $k_BT\gg W_\mathrm{t,c/v}$, the slope  is inversely proportional to $k_BT$. Therefore, the proposed relations correctly predict that the tail with the largest decay in the bandgap (either $k_BT$ of the Fermi-Dirac tail or $W_t$ of the band tail) dictates the slope of the charge-carrier density, which is consistent with Ref.\cite{limit} which used the Gauss hypergeometric function.

\section{Conclusion}
A closed-form expression is derived for a carrier density in an exponential band tail [expression (\ref{eq:zerothorder})], unifying the different cases used so far in the literature over temperature ($T$) and band-tail parameter ($W_t$). This expression is only valid for a position of $E_F$ in the tail states due to the convergence criterion ($\vert z\vert <1$) on the hypergeometric series expansion that was used for simplifying the Gauss hypergeometric function $H_{2F1}$. The final expressions including the band carriers [expressions \ref{eq:boltzn}) and (\ref{eq:boltzp})] are generalizations of the Boltzmann relations for blurred band edges. In the limit of sharp band edges (low disorder) and/or high temperature ($k_BT\gg W_t$), the proposed relations fall back on the Boltzmann relations, because the standard exponential that depends on temperature dominates the exponential that depends on the band-tail parameter. In the limit of high disorder and/or low temperature ($k_BT\ll W_t$), the proposed relations give back the frequently used expression from Shur and Hack \cite{shur_physics_1984} ($k_BT<W_t$) and the step-function approximation ($k_BT\ll W_t$). We have assumed a bulk density of states, but the given derivation can also apply to e.g., a constant 2-D DOS with a band tail\cite{pinsook_description_2013} provided some minor modifications are made. Identity (\ref{eq:zerothorder}) is not limited to charge-carrier densities, but applies more generally to any integral that can be brought into the form of an $H_{2F1}$ with that combination of arguments. The proposed relations can be used as a foundation for deriving analytical semiconductor-device models that have to include band tails, in particular models for MOSFETs at cryogenic temperatures or thin-film transistors. In addition, these relations will be useful in many simulation tools for materials in which band tails play a role.
\begin{figure*}[t]
	\includegraphics[width=0.95\textwidth]{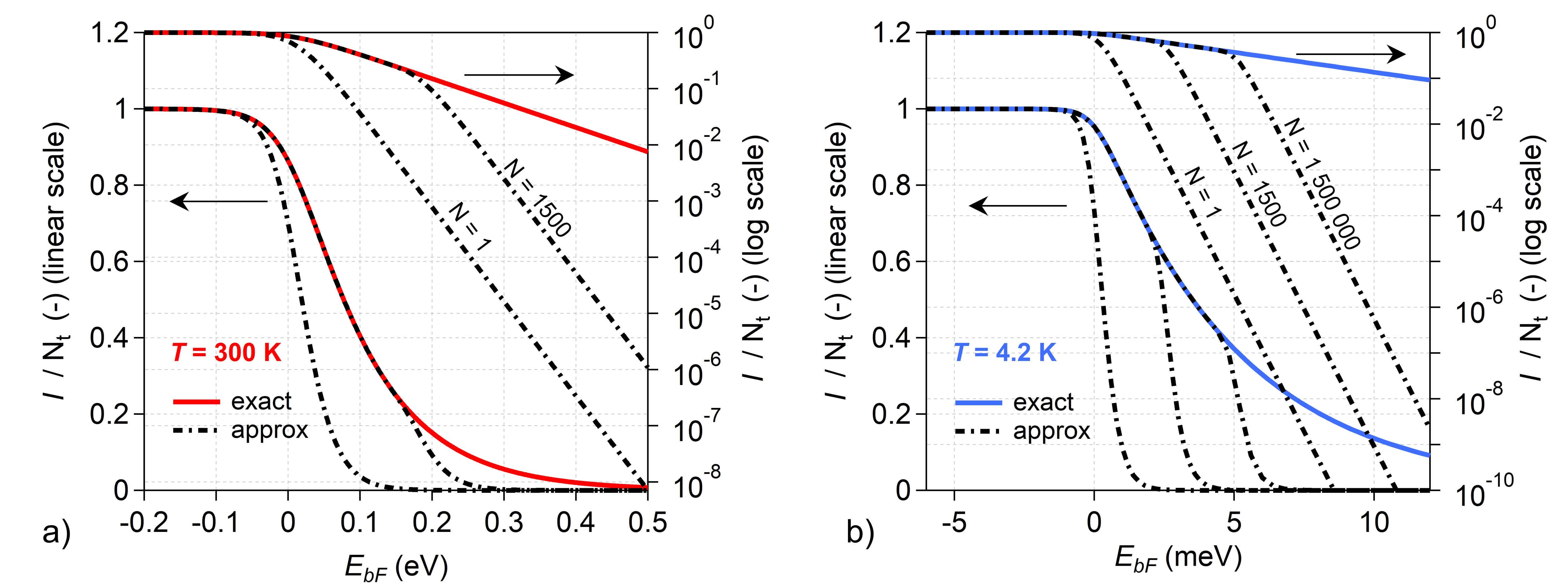}
	\caption{Exact expression (\ref{eq:nbt2}) versus approximate expression (\ref{eq:pfaffnbt}). Expression (\ref{eq:pfaffnbt}) has too slow convergence (too large $N$ required for $E_{bF}>0$) to be practically useful in analytical models. Examples: a) $W_t=\SI{100}{\milli\electronvolt}$ at RT, and b) $W_t=\SI{5}{\milli\electronvolt}$ at liquid-helium temperature.}
	\label{fig:pfaff}
\end{figure*}

\begin{acknowledgments}
This work was supported by the EU H2020 SEQUENCE Project under grant agreement number 871764. A.B. would like to thank Dr. Joren Vanherck, Philippe Roussel, Dr. Alexander Grill, and Dr. Subramanian
Narasimhamoorthy (imec, Belgium) for helpful discussions.
\end{acknowledgments}

\section*{Data Availability}
The data that support the findings of this study are available from the corresponding author
upon reasonable request.

\appendix
\section{\label{app:gauss}Properties of the Gauss hypergeometric function}
The Gauss hypergeometric function is often written as $\prescript{}{2}{F_1}(a,b;c;z)$, where the 2 and 1 indicate the number of Pochhammer symbols appearing in the numerator and denominator of the hypergeometric series (\ref{eq:pow}).\cite{oldham_atlas_2009} In this article, we have used the notation $H_{2F1}$ to avoid confusion with the common notation of Fermi-Dirac integrals. A varied punctuation, i.e., $\prescript{}{2}{F_1}(a,b;c;x)$, is normally used to indicate that $a$ and $b$ can be swapped, i.e., $\prescript{}{2}{F_1}(b,a,c,x) = \prescript{}{2}{F_1}(a,b,c,x)$.\cite{oldham_atlas_2009} Below follow some useful properties for $H_{2F1}$:\cite{rhyzik,nist} 
\begin{enumerate}
	\item Integration: $\int H_{2F1}(a,b;c;z)dz$
	\begin{equation}
	=\frac{c-1}{(a-1)(b-1)}H_{2F1}(a-1,b-1;c-1;z)
	\label{eq:int}
	\end{equation} 
	\item Integral property (property not listed in standard references of special functions\cite{rhyzik,nist}; but found online\cite{wolfram}) : 
	\begin{equation}
	\int \frac{dz}{1+z^\alpha}=zH_{2F1}(1,1/\alpha;1+1/\alpha;-z^\alpha)
	\label{eq:fd}
	\end{equation} 
	\item Special value: \begin{equation}
	H_{2F1}(a,b,c,0)=1
	\label{eq:specialvalue}
	\end{equation}
	\item Power series expansion for $\vert z\vert <1$: 
	\begin{equation}
	H_{2F1}(a,b;c;z)=\sum_{j=0}^{N} \frac{(a)_j(b)_j}{(c)_j}\frac{z^j}{j!}
	\label{eq:pow}
	\end{equation}
	where $()_j$ stands for the Pochhammer symbol 
	\begin{equation}
	(q)_j=q(q+1)\cdots(q+j-1),
	\end{equation} 
	and $(q)_0=1$.
	\item Pfaff transformations: $H_{2F1}(a,b;c;z)$
	\begin{subequations}
		\begin{eqnarray}	&=&(1-z)^{-a}H_{2F1}\left(a,c-b;c;\frac{z}{z-1}\right)
		\label{eq:app-pfaff}\\
		\label{eq:pfaff}
		&=&(1-z)^{-b}H_{2F1}\left(b,c-a;c;\frac{z}{z-1}\right)
		\end{eqnarray}
	\end{subequations}
	\item Linear combination transformation [9.132 in Ref.\cite{rhyzik}, 8th edition]: $H_{2F1}(a,b;c;z)$
	\begin{eqnarray}
	\label{eq:app-lco}
	=&\zeta^a&\frac{\Gamma (c)\Gamma(b-a)}{\Gamma(b)\Gamma(c-a)}H_{2F1}\left(a,c-b;a-b+1;\zeta\right)\\
	&+&\zeta^b\frac{\Gamma (c)\Gamma(a-b)}{\Gamma(a)\Gamma(c-b)}H_{2F1}\left(b,c-a;b-a+1;\zeta\right)\nonumber
	\end{eqnarray}
	where $\zeta=1/(1-z)$, $\Gamma$ the Gamma function, and $a-b \neq \pm m$ with $m\in \mathbb{N}$.
\end{enumerate}
\section{\label{sec:pfaff}Pfaff transformation}
The simplest of $H_{2F1}$-transformations available (either one of the Pfaff transformations, [(\ref{eq:app-pfaff}) or (\ref{eq:pfaff})] could be used in principle on (\ref{eq:nbt2}), since $\vert\zeta\vert=\vert z/(z-1)\vert<1$ for $E_F$ in the tail states. Applying this transformation to $H_{2F1}$ in (\ref{eq:nbt2}), gives: 
\begin{equation}
H_{2F1}\left(1,\theta;\theta+1;z\right)=(1-z)^{-1}\sum_{j=0}^{N} \frac{(1)_j(1)_j}{(\theta+1)_j}\frac{\zeta^j}{j!}.
\label{eq:pfaffnbt}
\end{equation}
However, the convergence of this series is slow since $\vert \zeta\vert \approx 1$. This can be understood from Fig.\,\ref{fig:pfaff} where a large number of terms ($N$) is still not sufficient to approximate the exact expression (\ref{eq:nbt2}). Moreover, the large $N$ required in the Pochhammer symbols and factorial causes arithmetic overflow, forcing us to work with multiprecision arithmetic even at RT. Therefore, convergence of (\ref{eq:pfaffnbt}) is too slow to be practically useful.

\section{\label{sec:higher}Higher-order approximations ($N=1,2$)}
Inserting $N=1$ and $N=2$ in (\ref{eq:S1}) and (\ref{eq:S2}), we obtain
\begin{subequations}
	\begin{eqnarray}
	\left(\frac{I}{N_t}\right)_{N=1}&=&\left(\frac{I}{N_t}\right)_{N=0}+\frac{A}{2-\theta}\exp\left(\frac{-2E_{bF}}{k_BT}\right)\nonumber\\&+&B\theta\exp\left(\frac{-E_{bF}(\theta+1)}{k_BT}\right)
	\label{eq:N1}\\
	\left(\frac{I}{N_t}\right)_{N=2}&=&\left(\frac{I}{N_t}\right)_{N=1}+\frac{2A}{(2-\theta)(3-\theta)}\exp\left(\frac{-3E_{bF}}{k_BT}\right)\nonumber\\&+&B\frac{\theta(\theta+1)}{2}\exp\left(\frac{-E_{bF}(\theta+2)}{k_BT}\right)
	\label{eq:N2}
	\end{eqnarray}
\end{subequations}

\section{\label{sec:app}Fermi-Dirac integral of band carriers}
Integral (\ref{eq:nb}),
\begin{equation}
n_b=\int_{E^*_c}^{\infty}C_e\sqrt{E-E_c}f(E)dE,
\end{equation}
has a different integration boundary ($E_c^*$) than the standard FD integral which starts at $E_c$. Using the Maxwell-Boltzmann approximation and the substitution $x=(E-E_c)/(k_BT)$, gives
\begin{equation}
n_b=C_e(k_BT)^{3/2}\exp\left(\frac{E_{F,n}-E_c}{k_BT}\right)\int_{1/(2\theta_c)}^{\infty}e^{-x}\sqrt{x}dx,
\end{equation}
where 
\begin{subequations}
	\begin{eqnarray}
	\int_{1/(2\theta_c)}^{\infty}e^{-x}\sqrt{x}dx&=&\int_{0}^{\infty}e^{-x}\sqrt{x}dx-\int^{1/(2\theta_c)}_{0}e^{-x}\sqrt{x}dx\nonumber\\
	&=&\frac{\sqrt{\pi}}{2}-\gamma\left(\frac{3}{2},\frac{1}{2\theta_c}\right), 
	\end{eqnarray}
\end{subequations}
where $\gamma$ stands for the lower incomplete gamma function.\cite{rhyzik} We obtain
\begin{equation}
n_b=C_e(k_BT)^{3/2}\frac{\sqrt{\pi}}{2}\left[1-\frac{2}{\sqrt{\pi}}\gamma\left(\frac{3}{2},\frac{1}{2\theta_c}\right)\right]\exp\left(\frac{E_{F,n}-E_c}{k_BT}\right)
\end{equation}
and since $N_c=C_e(k_BT)^{3/2}\sqrt{\pi}/2$, this yields
\begin{equation}
n_b=N_c^\prime\exp\left(\frac{E_{F,n}-E_c}{k_BT}\right),
\end{equation}
where 
\begin{equation}
N_c^\prime=N_c\left[1-\frac{2}{\sqrt{\pi}}\gamma\left(\frac{3}{2},\frac{1}{2\theta_c}\right)\right].
\end{equation}
Similarly for the hole density in the valence band in (\ref{eq:pb}), we obtain
\begin{equation}
p_b=N_v^\prime\exp\left(\frac{E_v-E_{F,p}}{k_BT}\right),
\end{equation}
where 
\begin{equation}
N_v^\prime=N_v\left[1-\frac{2}{\sqrt{\pi}}\gamma\left(\frac{3}{2},\frac{1}{2\theta_v}\right)\right]. 
\end{equation}

\bibliography{references}

\end{document}